\documentclass[aps,prl,twocolumn,superscriptaddress,
floatfix]{revtex4}
\usepackage{graphicx,epsfig}
\usepackage{amssymb,times}
\usepackage{amsbsy,amsmath,amsfonts,amsthm,bm}
\usepackage{color}
\usepackage{epstopdf}

\newcommand{\comment}[1]{{\bf #1}}

\def\be{\begin{equation}}
\def\ee{\end{equation}}

\def \rev@citealpnum#1{{\onlinecite{#1}}}

\def \comment#1{{}}

\def \subsection#1{{\bf #1.}}

\begin{document}
	
\title{Stability, Isolated Chaos, and Superdiffusion in Nonequilibrium Many-Body Interacting Systems}
\author{Atanu Rajak}
\affiliation{Department of Physics, Bar-Ilan University, Ramat Gan 5290002, Israel}
\affiliation{Presidency University, Kolkata, West Bengal 700073, India}
\author{Itzhack Dana}
\affiliation{Department of Physics, Bar-Ilan University, Ramat Gan 5290002, Israel}

\begin{abstract}
We demonstrate that stability and chaotic-transport features of paradigmatic nonequilibrium many-body systems, i.e., periodically kicked and interacting particles, can deviate significantly from the expected ones of full instability and normal chaotic diffusion for arbitrarily strong chaos, arbitrary number of particles, and different interaction cases. We rigorously show that under the latter general conditions there exist {\em fully stable} orbits, accelerator-mode (AM) fixed points, performing ballistic motion in momentum. It is numerically shown that an {\em ``isolated chaotic zone"} (ICZ), separated from the rest of the chaotic phase space, remains localized around an AM fixed point for long times even when this point is partially stable in only a few phase-space directions and despite the fact that Kolmogorov-Arnol'd-Moser tori are not isolating. The time evolution of the mean kinetic energy of an initial ensemble containing an ICZ exhibits {\em superdiffusion} instead of normal chaotic diffusion.
\end{abstract}

\maketitle

There has been a considerable interest, especially in the recent years, in understanding the classical and quantum properties of nonequilibrium many-body Hamiltonian systems given by periodically driven interacting particles \cite{kaneko89diffusion,konishi90diffusion,falcioni91ergodic,chirikov1993theory,chirikov97arnold,mulansky11strong,
rajak2018stability,nirfsr,rddt,gritsev2017integrable,ponte15many,lazarides15fate,ponte15periodically,abanin16theory,
agarwal17localization,dumitrescu2017logarithmically,choudhury14stability,bukov15prethermal,citro2015dynamical,
goldman15periodically,chandran16interaction,lellouch17parametric,lellouch2018parametric,abanin15exponentially,
kuwahara16floquet,mori2016rigorous,abanin17rigorous,howell2019asymptotic,mori2018floquet}. The periodic drive of a closed many-body system generically leads to heating. In well-known classical systems, this heating manifests itself in an unbounded chaotic diffusion of the total kinetic energy asymptotically in time \cite{kaneko89diffusion,konishi90diffusion,falcioni91ergodic,chirikov1993theory,chirikov97arnold,mulansky11strong,
rajak2018stability,nirfsr,rddt}. This diffusion is expected to occur generically even for arbitrarily weak nonintegrability (chaos strength) if the number of degrees of freedom is sufficiently large \cite{via,chirikov79universal} (see also below). 

There are open and fundamental questions concerning classical stability and chaotic transport in periodically driven many-body systems with $N>1$ degrees of freedom, e.g., $N$ interacting particles in one dimension. These questions naturally arise when considering the case of $N=1$, corresponding to the simplest Hamiltonian systems with chaotic dynamics, a famous paradigm being the periodically kicked rotor \cite{chirikov79universal}. The Poincar\'{e} map \cite{note} of such relatively simple systems already features an intricate mixture of chaotic and regular motions on all scales of a two-dimensional (2D) phase space \cite{note1,Karney,MO}. Stable periodic orbits are surrounded by 2D stability islands, see Fig. 1, and seem to exist for arbitrarily strong nonintegrability \cite{UF}, see also note \cite{note2}. The island boundaries are one-dimensional Kolmogorov-Arnol'd-Moser (KAM) tori \cite{K,A,M,cm} which thus fully separate the islands from the 2D chaotic region, i.e., they form barriers to chaos. Chaotic orbits can only stick to the island boundaries for long times, leading to significant deviations of chaos from fully random motion, with a slow decay of correlations \cite{Karney,MO}. At the same time, the motion inside an island is essentially a regular one around the island center, a point of the stable periodic orbit to which the island is associated.

\begin{figure}[t]
\includegraphics[width=9cm,height=4.5cm]{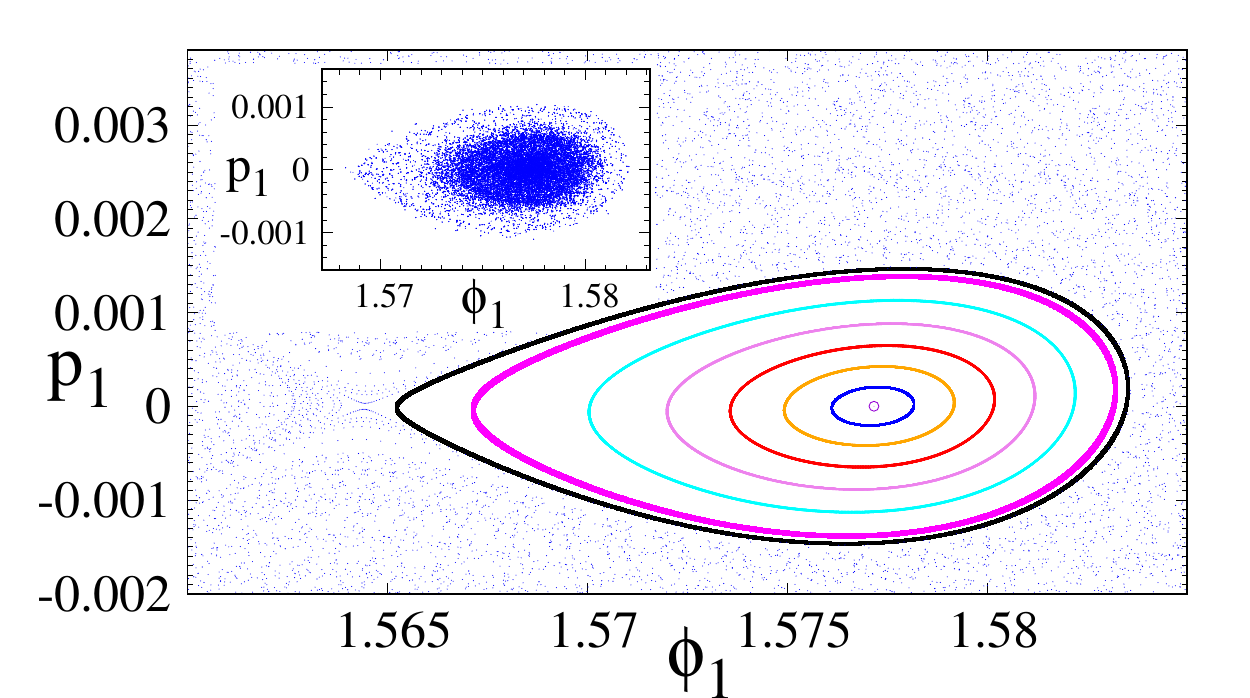}
\setlength{\abovecaptionskip}{-10pt}
\setlength{\belowcaptionskip}{-10pt}
\caption{Phase space of the standard map \{map (\ref{pm}) for $N=1$, with $p_1$ taken mod$[-\pi ,\pi )$\}, showing the chaotic region and a stability island around an accelerator-mode (AM) fixed point, defined by Eq. (\ref{fp}) with $w_1=1$ and for $\tau_1=1.96$ in Eq. (\ref{t1}) ($K\approx 6.2833$). Inset: Isolated chaotic zone for $N=4$ with interaction, analogous to the $N=1$ AM island above; see details in the text and in Fig. 2.} 
\label{fig1}
\end{figure}  

For $N>1$, on the other hand, KAM tori, in particular tori surrounding stable periodic orbits, are generically $N$-dimensional surfaces \cite{note3} which cannot be isolating barriers in the $2N$-dimensional phase space, due to purely topological reasons. Thus, chaos can now penetrate the interior of KAM tori by the so-called Arnol'd diffusion \cite{via,chirikov79universal} which is, however, extremely slow in the limit of vanishing nonintegrability strength \cite{NN}. One can then ask the following questions: Do stable periodic orbits exist even in cases where {\em one may expect a fully chaotic phase space}, i.e., for arbitrarily large $N$ and, like in the $N=1$ case, for arbitrarily strong nonintegrability? Are these orbits surrounded by ``stability regions", analogous to the stability islands for $N=1$ and leading to significant deviations of chaotic transport from normal diffusion? 

In this Letter, we answer these questions affirmatively for paradigmatic and realistic many-body systems, the answer to the first question being a rigorous one. We study periodically kicked systems described by the general Hamiltonian \cite{nirfsr,note4}
\begin{align}\label{eq:H}
H(t)&=\sum_{j=1}^N \left[ \frac{p_j^2}{2}+K\cos(\phi_j)\Delta(t)\right] \nonumber \\ 
&+\sum_{i,j=1,i\neq j}^N\eta_{i,j}\cos(\phi_i-\phi_j)\Delta(t)\;.
\end{align}
Here $p_j$ and $\phi_j$ ($j=1,...,N$) are, respectively, the (angular) momenta and positions (angles) of $N$ rotors (particles on a circle), $K$ is a parameter, $\Delta(t)=\sum_{s=-\infty}^{\infty} \delta(t-s)$ is a periodic delta function with time period $T=1$, and $\eta_{i,j}$ are interaction strengths. Two extreme cases of interactions will be studied: The infinite-range ones, $\eta_{i,j}=\eta /N$, and the nearest-neighbor ones, $\eta_{i,j}=\eta \delta_{i,j-1}$, where $\eta$ is a positive constant. In the nearest-neighbor case, the angles $\phi_j$ satisfy periodic boundary conditions, $\phi_{N+j}=\phi_j$, defining also $\phi_0=\phi_N$. In this case and for $K=0$, the Hamiltonian (\ref{eq:H}) describes paradigmatic model systems of many-body chaos theory \cite{kaneko89diffusion,konishi90diffusion,falcioni91ergodic,chirikov1993theory,chirikov97arnold,mulansky11strong,
rajak2018stability,nirfsr,rddt} that may be experimentally realizable (see, e.g., Ref. \cite{rddt}). The generalizations of these systems to $K\neq 0$ should be experimentally realizable as well \cite{pc}. The case of $N=2$ for $K\neq 0$ was investigated in several works \cite{F,FS,KM,RLBK,LROBK,OLKB}. Some properties of the quantum counterparts of the general systems (\ref{eq:H}) for both kinds of interactions were studied recently \cite{nirfsr}. 

We rigorously show that fully stable orbits of the Poincar\'{e} map for the systems (\ref{eq:H}), in both interaction cases, exist for nonintegrability strength $K$ unbounded from above and for arbitrary number $N$ of particles, despite of the expectation that the phase space should be completely unstable for large $K$ and $N$. The stable orbits are accelerator-mode (AM) fixed points that are linearly stable in all directions of the multidimensional phase space. The AM fixed points, well known in the single-particle ($N=1$) chaos theory \cite{chirikov79universal,am1,am2,am3,am4,am5,am6,am7,am8,am9,am10,am11,am12,am13,am14,am15,am16} [see definition (\ref{fp}) below], perform ballistic motion in momentum and provide the most well established mechanism of chaotic superdiffusion for $N=1$ \cite{am9,am10,am11,am12,am13,am14,am15,am16}, roughly as follows. An ensemble of chaotic orbits sticking to boundaries of AM stability islands performs ballistic motion in momentum, i.e., its mean kinetic energy increases quadratically in time like that of an initial ensemble inside an AM island. However, when the chaotic ensemble leaves the boundaries of AM islands, it performs normal diffusion in the chaotic sea, namely, its mean kinetic energy increases linearly in time. The average result of these processes over a very long time is superdiffusion of the chaotic ensemble, with its mean kinetic energy increasing in time between linearly and quadratically.    

While for $N>1$ KAM tori do not form barriers and stability islands do not strictly exist, we provide numerical evidence that, for an initial ensemble that is sufficiently localized around an AM fixed point, part of this ensemble remains localized there for a long time if the AM fixed point is linearly stable already just in a few phase-space directions. The localization takes place in a ``chaotic zone" (see inset of Fig. 1 and below) that is ``stable" in the sense that it remains {\em isolated} from the rest of the chaotic region for a long time. All the points in such a zone essentially perform ballistic motion in momentum. Thus, the existence of an isolated chaotic zone (ICZ) around an AM fixed point leads to a superdiffusion of the entire initial ensemble. The ICZ may be viewed as an $N>1$ analog of a regular stability island for $N=1$. 
  
We start by writing the Poincar\'{e} map from kick to kick for the systems (\ref{eq:H}). Denoting $p_j(s)=p_j(t=s-0)$ and $\phi_j(s)=\phi_j(t=s-0)$ ($s$ integer), one can easily derive from Hamilton equations the exact map equations
\begin{align}\label{pm}
p_j(s+1) &=p_j(s)+K\sin[\phi_j(s)]+\eta F_j[\phi_1(s),...,\phi_N(s)], \nonumber \\ \nonumber \\
\phi_j(s+1) &= \phi_j(s)+p_j(s+1)\ {\rm mod}(2\pi ),
\end{align}
where the mod($2\pi$) is taken since $\phi_j(s)$ is an angle,
\begin{equation}\label{fi}
F_j[\phi_1(s),...,\phi_N(s)]=\frac{1}{N}\sum_{i=1}^N\sin[\phi_j(s)-\phi_i(s)]
\end{equation}
in the infinite-range case and 
\begin{equation}\label{fnn}
F_j[\phi_1(s),...,\phi_N(s)]=\sum_{i=-1}^1 \sin[\phi_j(s)-\phi_i(s)] 
\end{equation}  
in the nearest-neighbor case.

The map (\ref{pm}) is clearly translationally invariant in both $p_j(s)$ and $\phi_j(s)$ with period $2\pi$. This invariance allows to define the fixed points of the map in a generalized way:
\begin{equation}\label{fp}
p_j(s+1) =p_j(s)+2\pi w_j,\ \ \ \phi_j(s+1) = \phi_j(s),
\end{equation}
where $w_j$ are arbitrary integers. For $w_j\neq 0$, Eq. (\ref{fp}) defines accelerator-mode (AM) fixed points performing ballistic motion in momentum with acceleration $2\pi w_j$. A natural exact solution for Eq. (\ref{fp}) can be obtained if the quantities $p_j(0)$, $w_j$, and $\phi_j(0)$, are independent of $j$, $p_j(0)=p$, $w_j=w$, and $\phi_j(0)=\phi$ for all $j$. Using Eqs. (\ref{pm})-(\ref{fp}), we then get
\begin{equation}\label{cam}
p\ {\rm mod}(2\pi )=0,\ \ \ K\sin(\phi)=2\pi w.
\end{equation}

We now study the linear stability of the AM fixed points (\ref{fp}) with (\ref{cam}) under small perturbations. This stability is determined by the $2N\times 2N$ derivative matrix $\mathbf{DM}$ of the map $\mathbf{M}$ in Eq. (\ref{pm}), where the derivatives are taken with respect to the $2N$ phase-space variables $(p_j,\phi_j)$, $i,j=1,...,N$, and are evaluated at the AM fixed points. After a straightforward calculation, we find that
\begin{equation}\label{DM}
\mathbf{DM}=
\begin{pmatrix}
{\mathbf I} & {\mathbf A} \\
{\mathbf I} & {\mathbf I}+{\mathbf A}
\end{pmatrix}
, 
\end{equation} 
where $\mathbf{I}$ is the $N\times N$ unit matrix and $\mathbf{A}$ is the $N\times N$ matrix with elements
\begin{equation}\label{aei}
\mathbf{A}_{i,j}= -\eta /N+[K\cos(\phi)+\eta ]\delta_{i,j}
\end{equation} 
in the infinite-range case and
\begin{equation}\label{aenn}
\mathbf{A}_{i,j}= -\eta (\delta_{i-1,j}+\delta_{i,j-1})+[K\cos(\phi)+2\eta ]\delta_{i,j}
\end{equation}
in the nearest-neighbor case, $i,j=1,...,N$, with periodic boundary conditions in Eq. (\ref{aenn}). Being the derivative matrix of a Hamiltonian map (\ref{pm}), the matrix (\ref{DM}) is symplectic (see, e.g., Ref. \cite{rmk}); in particular, its $2N$ eigenvalues form $N$ reciprocal pairs $(\lambda_j,\lambda_j^{-1})$, $j=1,...,N$. One has linear stability in all $2N$ phase-space directions only if all the eigenvalues lie on the unit circle in the complex plane, $|\lambda_j|=1$, $j=1,...,N$. Defining $\tau_j=\lambda_j+\lambda_j^{-1}$, the full stability condition is thus
\begin{equation}\label{sc}
|\tau_j|<2,\ \ j=1,...,N.
\end{equation} 

A main result of this Letter are exact expressions for $\tau_j$ in both interaction cases. The details of the derivation of these expressions are given in the Supplemental Material below. In the infinite-range case, we find that
\begin{equation}\label{ti}
\tau_1=2+K\cos(\phi ),\ \ \ \ \tau_j=\tau_1+\eta
\end{equation}
for $j=2,...,N$ and, from Eq. (\ref{cam}), one can write
\begin{equation}\label{t1}
\tau_1=2+K\cos(\phi )=2\pm\sqrt{K^2-4\pi^2w^2}.
\end{equation}
In the nearest-neighbor case, we obtain, for $j=1,...,N$, 
\begin{equation}\label{tnne}
\tau_j=2+K\cos(\phi )+4\eta \cos ^2(\pi j/N)
\end{equation}
if $N$ is even and 
\begin{equation}\label{tnno}
\tau_j=2+K\cos(\phi )+4\eta \cos ^2[\pi (j+0.5)/N]
\end{equation}
if $N$ is odd. It is clear from Eqs. (\ref{cam}) and (\ref{t1}) that one can ensure that $|2+K\cos(\phi )|<2$ for $K$ unbounded from above by choosing $w$ sufficiently large, so that 
\begin{equation}\label{KB}
2\pi w< K< \sqrt{16+4\pi^2w^2} .
\end{equation} 
Then, provided $\eta$ is sufficiently small, Eqs. (\ref{ti}), (\ref{tnne}), and (\ref{tnno}) imply the full stability condition (\ref{sc}) for $K$ unbounded from above and for arbitrarily large $N$.

We now consider the implications of the stability conditions (\ref{sc}), satisfied at least for some $j$ values, on the stability in the neighborhood of an AM fixed point and on the chaotic transport. In our numerical calculations, the initial ensemble $(\phi_j(0),p_j(0))$, $j=1,...,N$, is a $2N$-dimensional grid in a small hypercube of side $2\epsilon$ around the AM fixed point: $|\phi_j(0)-\phi |\leq \epsilon$, $|p_j(0)-p|\leq \epsilon$, where $(\phi ,p)$ are given by Eq. (\ref{cam}). If the number of grid points on each side of the hypercube is $L$, the number of points in the ensemble is $L^{2N}$. Unless otherwise specified, we assume in what follows for definiteness the infinite-range case and the values $w=1$, $N=4$, $L=5$. We have checked that our results hold in other cases, including the nearest-neighbor interaction case. 

Figure 2 shows the time evolution of the initial ensemble above for $\epsilon =0.001$, $\tau_1=1.96$, and $\eta =0.01$. From Eq. (\ref{ti}), it is clear that in this case the conditions (\ref{sc}) are satisfied for all $j$, i.e., the AM fixed point is fully stable. We see a gradual depletion of the ensemble as the number of iterations $s$ increases, $s=30,50$, with points escaping in the left $\phi$ direction. However, for sufficiently large $s\gtrsim 1000$, the escape essentially stops and one is left with an ICZ which remains well localized around the AM fixed point at least until $s=10^5$, see Figs. 2(c), 2(d), and the inset of Fig. 1.

\begin{figure}[t]
\includegraphics[width=9cm,height=8cm]{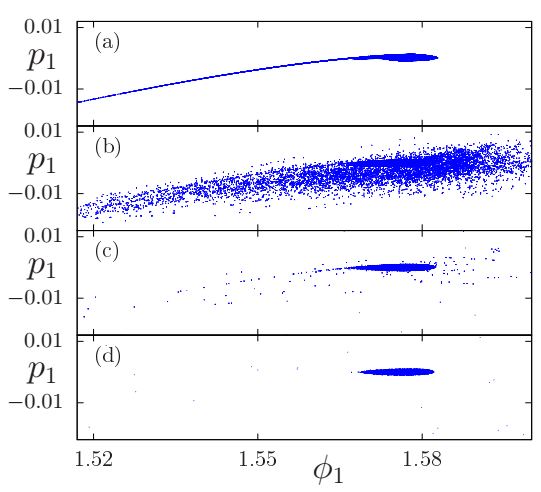}
\setlength{\abovecaptionskip}{-10pt}
\setlength{\belowcaptionskip}{-7pt}
\caption{Formation of an ICZ from the time evolution [projected in the $(\phi_1,p_1)$ phase plane] of the initial ensemble described in the text ($\epsilon =0.001$) under the map (\ref{pm}) for $N=4$ [with $p_j$ taken mod$[-\pi ,\pi )$], $\eta =0.01$ (infinite-range case), $w=1$, $\tau_1=1.96$ in Eq. (\ref{t1}) ($K\approx 6.2833$), and number of iterations: (a) $s=30$; (b) $s=50$; (c) $s=1000$; (d) $s=10^5$. The inset in Fig. 1 shows a zoom of the ICZ clearly emerging in case (d). The number of points in this ICZ is $\sim 58130$, out of $5^8=390625$ points in the initial ensemble.} 
\label{fig2}
\end{figure}

Figure 3 shows that an ICZ continues to exist even for larger values of the interaction strength, such as $\eta =1$. This despite the fact that for this value of $\eta$ the AM fixed point is only partially stable according to Eq. (\ref{ti}), with $|\tau_1|<2$ but $|\tau_j|>2$ for $j=2,3,4$. By increasing the number of points in the initial ensemble, the number of points in the ICZ increases, compare Fig. 3(a) with Fig. 3(b). 

\begin{figure}[t]
\includegraphics[width=9cm,height=4.5cm]{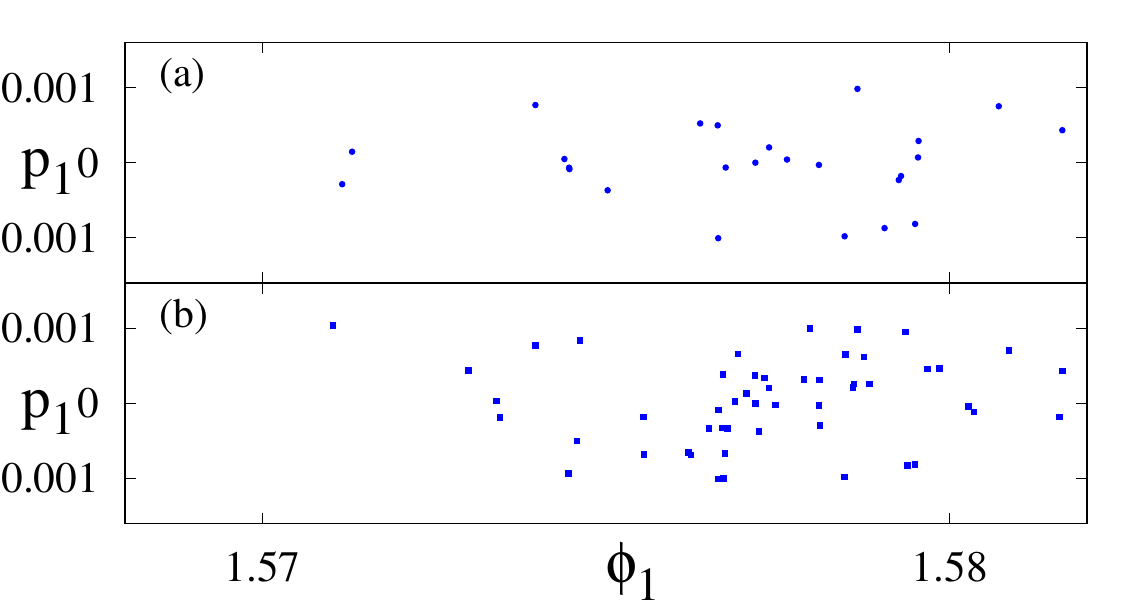}
\setlength{\abovecaptionskip}{-10pt}
\setlength{\belowcaptionskip}{-7pt}
\caption{ICZ formed after $s=10^5$ map iterations for the same parameter values as in Fig. 2, except that $\eta =1$ and, in case (b), the initial ensemble has $7^8$ points (with 51 points in the ICZ) instead of $5^8$ points in case (a) (with 24 points in the ICZ).}
\label{fig3}
\end{figure}

Being well localized around an AM fixed point for a long time, an ICZ behaves like an AM stability island for $N=1$ (see Fig. 1); namely, from Eq. (\ref{fp}) it follows that the mean kinetic energy $E_{\rm K}(s)=\left\langle\sum_{j=1}^N p_j^2(s)/2\right\rangle$, with the average taken over the ICZ as initial ensemble, increases precisely as $s^2$ (quadratically). We have explicitly verified this in several examples, including those in Figs. 2 and 3. 

However, if the average in $E_{\rm K}(s)$ is taken over the entire initial ensemble, $E_{\rm K}(s)$ will increase slower than $s^2$ since most of the points in the initial ensemble are outside the ICZ. Thus, the contribution of these points to       $E_{\rm K}(s)$ should be almost diffusive, i.e., $\sim s^{\mu }$ with $\mu \sim 1$. As $s$ increases, the contribution of the ICZ to $E_{\rm K}(s)$ will become more and more dominant, even for arbitrarily small number of points in the ICZ. Therefore, we expect that the average slope AS$(s)$ of the graph of $\ln [E_{\rm K}(s)]$ versus $\ln (s)$ in a time interval $[s_0,s]$, for some initial time $s_0$, will satisfy AS$(s)>1$ (superdiffusion) for sufficiently large $s$. This is shown in Fig. 4 for several values of $\eta$. We found that when the AM fixed point is partially stable, e.g., $\tau_1<2$ and $\tau_1+\eta>2$, the value of AS$(s)>1$ for large $s$ seems to be almost independent of $\eta$, as shown in Fig. 4 for $\eta=0.05$ and $\eta =1$; this indicates that the size of the ICZ does not almost change in some interval of relatively large values of $\eta$.

\begin{figure}[t]
\includegraphics[width=9cm,height=4.5cm]{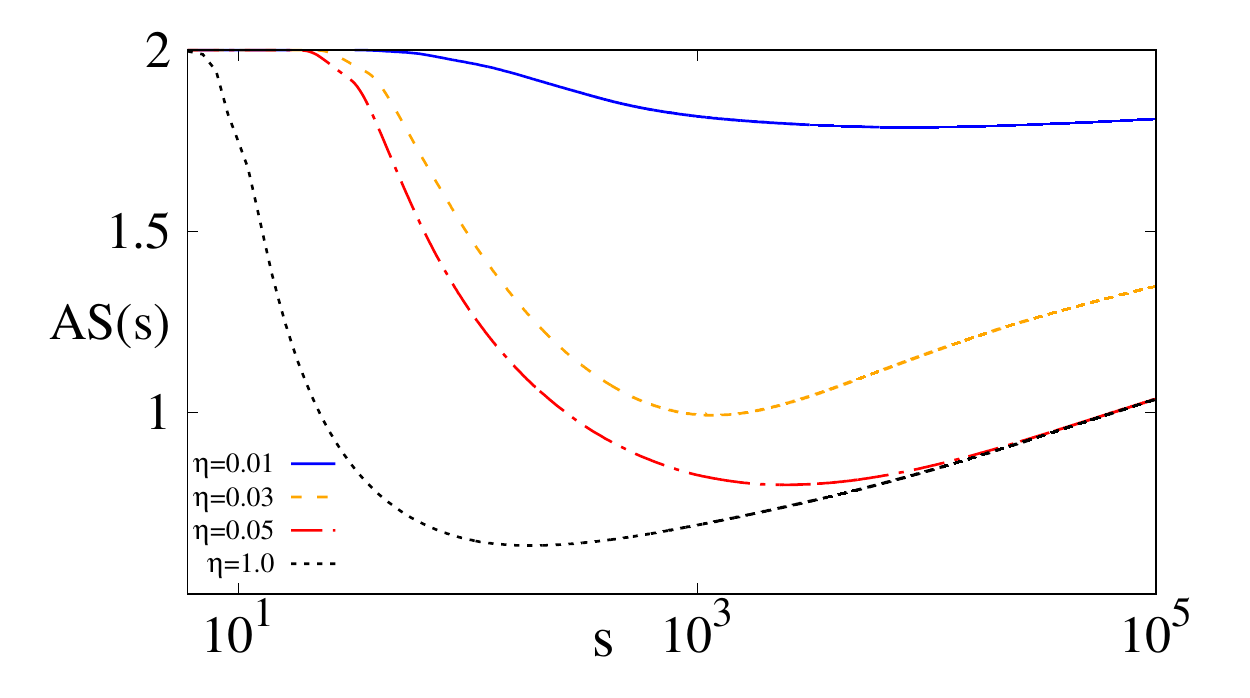}
\setlength{\abovecaptionskip}{-10pt}
\setlength{\belowcaptionskip}{-7pt}
\caption{Average slope AS$(s)$ of the graph of $\ln [E_{\rm K}(s)]$ versus $\ln (s)$, in a time interval $[s_0,s]$, $s_0=5$, for $N=4$, $w=1$, $\tau_1=1.96$, and several values of $\eta$.}
\label{fig4}
\end{figure}

When the AM fixed point is fully unstable, e.g., when $\tau_1>2$, no ICZ exists and no superdiffusion is observed: AS$(s)\leq 1$ for all $\eta$. Figure 5 shows that this is approximately the case also under conditions of almost instability, e.g., when $\tau_1\lesssim 2$ ($\tau_1=1.999$ in Fig. 5), even for very small $\eta$.

\begin{figure}[t]
\includegraphics[width=9cm,height=4.5cm]{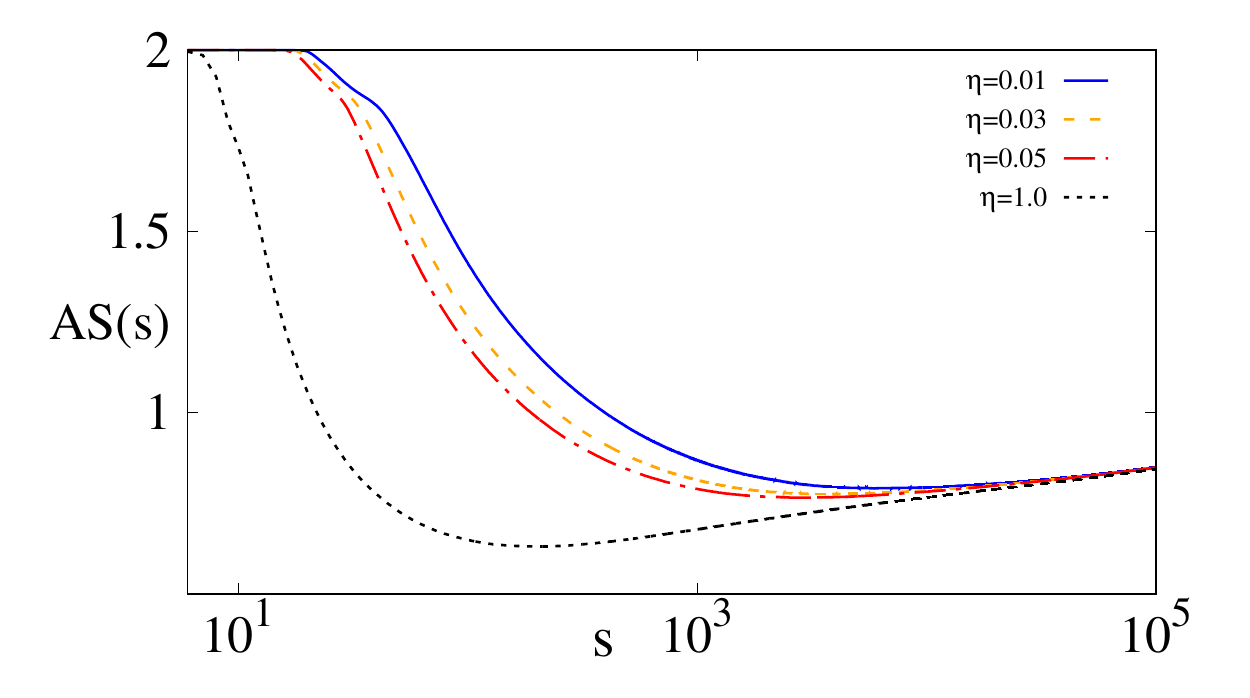}
\setlength{\abovecaptionskip}{-10pt}
\setlength{\belowcaptionskip}{-7pt}
\caption{Similar to Fig. 4 but for $\tau_1=1.999$.}
\label{fig5}
\end{figure}

In conclusion, we have rigorously shown that fully stable orbits exist in the paradigmatic nonequilibrium many-body systems (\ref{eq:H}) in two extreme cases of interactions and under conditions where one would expect a completely unstable and chaotic phase space: Arbitrary number $N$ of particles and nonintegrability (chaos) strength $K$ unbounded from above. 

We have provided numerical evidence for $N>1$ that even when the considered orbits (AM fixed points) are partially stable, i.e., stable not in all phase-space directions, the orbits are surrounded by an ICZ, separated from the rest of the chaotic phase space and analogous to a stability island for $N=1$; see Figs. 1-3 and compare Fig. 1 with its inset. The existence of ICZs, especially under partial stability conditions, is surprising in view of the fact that KAM tori do not form strict barriers for $N>1$, unlike the case of $N=1$ where a stability island is impenetrable from the chaotic region.

All the points in an ICZ perform ballistic motion in momentum like the AM fixed point to which the ICZ is associated. This motion causes significant deviations of the chaotic transport from the normal diffusion featured by a completely chaotic system. Namely, the mean kinetic energy of an ensemble containing an ICZ exhibits superdiffusion even for relatively large interaction strength, see Fig. 4. 

The existence of ICZs may be understood as due to trapping of chaotic orbits inside a set of many KAM tori surrounding a fixed point. Such a set may form a very effective barrier to chaotic transport in the vicinity of fixed points. We plan to investigate how ICZs emerge in detail in future works.

\begin{center}
{\bf Supplemental Material}
\end{center} 

\renewcommand{\theequation}{SM\arabic{equation}}
\setcounter{equation}{0}

We derive here the exact results (11)-(14) in the paper, in the two interaction cases. In both cases, the eigenvalues $\lambda=\lambda_j$, $j=1,...,2N$, of the matrix $\mathbf{DM}$ [Eq. (7) in the paper] are determined from $\det(\mathbf{B}_{\lambda})=0$, where
\begin{equation}\label{Bl}
\mathbf{B}_{\lambda}=\mathbf{DM}-\lambda\mathbf{I}=
\begin{pmatrix}
(1-\lambda ){\mathbf I} & {\mathbf A} \\
{\mathbf I} & (1-\lambda ){\mathbf I}+{\mathbf A} 
\end{pmatrix}
.
\end{equation}
For $j\leq N$, let us subtract row $j+N$ from row $j$ in Eq. (\ref{Bl}). This will transform $\mathbf{B}_{\lambda}$ into a matrix having the same determinant as $\mathbf{B}_{\lambda}$:
\begin{equation}\label{Blt}
\mathbf{B}'_{\lambda}=
\begin{pmatrix}
(1-\lambda ){\mathbf I} & {\mathbf A} \\
\lambda {\mathbf I} & (1-\lambda ){\mathbf I} 
\end{pmatrix}
.
\end{equation}

The matrix (\ref{Blt}) consists of four $N\times N$ blocks or sub-matrices that clearly commute with each other. Therefore, one can write: 
\begin{align}\label{dBl}
\det(\mathbf{B}_{\lambda})=\det(\mathbf{B}'_{\lambda}) &=\det[(1-\lambda )^2{\mathbf I}-\lambda {\mathbf A}] \nonumber \\
&=\det(\lambda \mathbf{A}_{\lambda ,N}),
\end{align}
where we defined the $N\times N$ matrix
\begin{equation}\label{Al}
\mathbf{A}_{\lambda ,N}=\left(\lambda +\lambda^{-1}-2\right){\mathbf I}-{\mathbf A}.
\end{equation} 
We now derive exact expressions for $\det(\mathbf{A}_{\lambda ,N})$ in the two interaction cases.

\begin{center}{\bf A. Infinite-interaction case}\end{center}

From Eq. (8) in the paper, the elements of the matrix (\ref{Al}) in this case are
\begin{equation}\label{alei}
(\mathbf{A}_{\lambda ,N})_{i,j}=\eta /N + a_{\lambda}\delta_{i,j} ,
\end{equation}
where 
\begin{equation}\label{al}
a_{\lambda}=\lambda +\lambda^{-1}-2-K\cos(\phi)-\eta .
\end{equation}
By subtracting the second row from the first row of $\mathbf{A}_{\lambda ,N}$, using Eq. (\ref{alei}), and expanding the determinant of the resulting matrix from its first row, we get:
\begin{equation}\label{edali}
\det(\mathbf{A}_{\lambda ,N}) =a_{\lambda}[\det(\mathbf{A}_{\lambda ,N-1})+\det(\mathbf{A}_{\lambda ,\eta ,N-1}) ] ,
\end{equation}
where $\mathbf{A}_{\lambda ,\eta ,N}$ is the matrix whose elements are 
\begin{equation}\label{aleei}
(\mathbf{A}_{\lambda ,\eta ,N})_{i,j}=(\mathbf{A}_{\lambda ,N})_{i,j} - a_{\lambda}\delta_{i,1}\delta_{j,1} 
\end{equation}
and $(\mathbf{A}_{\lambda ,N})_{i,j}$ are given by Eq. (\ref{alei}). By expanding the determinant of $\mathbf{A}_{\lambda ,\eta ,N}$ from its second column, using Eq. (\ref{aleei}), we find that
\begin{equation}\label{ralei}
\det(\mathbf{A}_{\lambda ,\eta ,N})=a_{\lambda}\det(\mathbf{A}_{\lambda ,\eta ,N-1}).
\end{equation}
Using the recursions in Eqs. (\ref{edali}) and (\ref{ralei}) repetitively, we obtain
\begin{align}\label{edali2}
\det(\mathbf{A}_{\lambda ,N}) &= a_{\lambda}^2[\det(\mathbf{A}_{\lambda ,N-2})+2\det(\mathbf{A}_{\lambda ,\eta ,N-2}) ] \nonumber \\
&=a_{\lambda}^3 [\det(\mathbf{A}_{\lambda ,N-3})+3\det(\mathbf{A}_{\lambda ,\eta ,N-3}) ] \nonumber \\
&=... \nonumber \\
&=a_{\lambda}^{N-2} [\det(\mathbf{A}_{\lambda ,2})+(N-2)\det(\mathbf{A}_{\lambda ,\eta ,2}) ] \nonumber \\
&=a_{\lambda}^{N-2} \left[\det\begin{pmatrix}
a_{\lambda}+\eta /N & \eta /N \\
\eta /N & a_{\lambda}+\eta /N 
\end{pmatrix} \right. \nonumber \\
&\left. +(N-2)\det\begin{pmatrix}
\eta /N & \eta /N \\
\eta /N & a_{\lambda}+\eta /N 
\end{pmatrix} \right] \nonumber \\
&=a_{\lambda}^{N-1}(a_{\lambda}+\eta ) .
\end{align}

From the eigenvalue equation $\det(\mathbf{B}_{\lambda})=0$ or, by Eq. (\ref{dBl}), $\det(\mathbf{A}_{\lambda ,N})=0$ (because $\lambda\neq 0$ as shown below), we get from Eqs. (\ref{al}) and (\ref{edali2}) that the $N$ solutions for $\tau_j=\lambda_j +\lambda_j^{-1}$, $j=1,...,N$, are given precisely by Eq. (11) in the paper. Since all these solutions are finite, $\lambda_j\neq 0$.

\begin{center}{\bf B. Nearest-neighbor-interaction case}\end{center}

From Eq. (9) in the paper, the elements of the matrix (\ref{Al}) in this case are
\begin{equation}\label{alenn}
(\mathbf{A}_{\lambda ,N})_{i,j}=\eta (\delta_{i-1,j}+\delta_{i,j-1}) + b_{\lambda}\delta_{i,j} ,
\end{equation}
where 
\begin{equation}\label{bl}
b_{\lambda}=\lambda +\lambda^{-1}-2-K\cos(\phi)-2\eta .
\end{equation}
The matrix (\ref{alenn}) is a tridiagonal one with periodic boundary conditions. Using known formulas for the determinant of such a matrix (see, e.g., Ref. \cite{lgm}), we get
\begin{equation}\label{dalnn}
\det(\mathbf{A}_{\lambda ,N})= -2(-\eta )^N +{\rm Tr}\left[ 
\begin{pmatrix} b_{\lambda} &  -\eta ^2\\ 1 & 0 \end{pmatrix} ^N \right].
\end{equation}
The trace of the $N$th power of the $2\times 2$ matrix in Eq. (\ref{dalnn}) can be exactly evaluated by first calculating the eigenvalues of this matrix:
\begin{equation}\label{gpm}
\gamma_{\pm }=\frac{b_{\lambda}}{2}\pm \sqrt{\left(\frac{b_{\lambda}}{2}\right)^2-\eta^2}.
\end{equation}
The trace in Eq. (\ref{dalnn}) is then given by $\gamma_{+}^N+\gamma_{-}^N=\gamma_{+}^N+\eta^{2N}\gamma_{+}^{-N}$. Using this in Eq. (\ref{dalnn}) and denoting $\gamma_{+}^N$ by $z$, the equation $\det(\mathbf{A}_{\lambda ,N})=0$ for the eigenvalues $\lambda_j$ will read:
\begin{equation}\label{eenn}
z^2-2(-\eta )^Nz+\eta^{2N}=\left[ z-(-\eta )^N\right]^2=0.
\end{equation}
Thus, $z=\gamma_{+}^N=(-\eta )^N$, with $N$ solutions for $\gamma_{+}$:
\begin{equation}\label{ge}
\gamma_{+,j}=\eta \exp \left[ \frac{i\pi (2j+\beta )}{N}\right] , 
\end{equation}
$j=1,...,N$, where $\beta =0$ for $N$ even and $\beta =1$ for $N$ odd. Using Eq. (\ref{ge}) in Eq. (\ref{gpm}), defining $\gamma_{+}$, and solving for $b_{\lambda}$, we get
\begin{equation}\label{ble}
b_{\lambda,j}=2\eta \cos \left[ \frac{i\pi (2j+\beta )}{N}\right] .
\end{equation}
Finally, using the definition (\ref{bl}) of $b_{\lambda}$ in Eq. (\ref{ble}), we obtain the $N$ values of $\tau_j=\lambda _j +\lambda _j^{-1}$, as given by Eqs. (13) and (14) in the paper.     

\begin{acknowledgments} 
We thank E. G. D. Torre for useful discussions. This work is supported by the Israel Science Foundation, grant no. 151/19.
\end{acknowledgments}


\begin{thebibliography}{38}
\bibitem[{\citenamefont{note}(2019)}]{note}
	\bibinfo{journal}{The Poincar\'{e} map of a periodically driven system, see, e.g., Eq. (\ref{pm}), is $2N$-
	dimensional. On the other hand, the Poincar\'{e} map of a time-independent Hamiltonian system with $N$ degrees of 
	freedom on the energy surface is $(2N-2)$-dimensional.}
\bibitem[{\citenamefont{note1}(2019)}]{note1}
	\bibinfo{journal}{Essentially the same phase-space structure is featured by the 2D Poincar\'{e} map of a 
	time-independent Hamiltonian system with two degrees of freedom on the energy surface. See note \cite{note}.}
\bibitem[{\citenamefont{note2}(2019)}]{note2}
	\bibinfo{journal}{In particular, the existence of stable accelerator-mode fixed points for nonintegrability strength       
	unbounded from above follows from our results below in the case of $N=1$, see sentence after Eq. (\ref{tnno}).}
\bibitem[{\citenamefont{note3}(2019)}]{note3}
	\bibinfo{journal}{In general, the KAM tori are surfaces of dimension $\leq N$; see, e.g., Ref. \cite{cm} and references 
	therein.}
\bibitem[{\citenamefont{note4}(2019)}]{note4}
	\bibinfo{journal}{Our notation for the Hamiltonian (\ref{eq:H}) and related quantities slightly differs from that in 
	Ref. \cite{nirfsr}.}
\end{thebibliography}
\end{document}